# Mining Compressed Repetitive Gapped Sequential Patterns Efficiently

Yongxin Tong, Li Zhao, Dan Yu, Shilong Ma, and Ke Xu

State Key Lab. of Software Development Environment
Beihang University, Beijing 100191, China
{yxtong, lzh, yudan, slma, kexu}@nlsde.buaa.edu.cn

**Abstract.** Mining frequent sequential patterns from sequence databases has been a central research topic in data mining and various efficient mining sequential patterns algorithms have been proposed and studied. Recently, in many problem domains (e.g, program execution traces), a novel sequential pattern mining research, called mining repetitive gapped sequential patterns, has attracted the attention of many researchers, considering not only the repetition of sequential pattern in different sequences but also the repetition within a sequence is more meaningful than the general sequential pattern mining which only captures occurrences in different sequences. However, the number of repetitive gapped sequential patterns generated by even these closed mining algorithms may be too large to understand for users, especially when support threshold is low.

In this paper, we propose and study the problem of compressing repetitive gapped sequential patterns. Inspired by the ideas of summarizing frequent itemsets, RPglobal, we develop an algorithm, CRGSgrow (Compressing Repetitive Gapped Sequential pattern grow), including an efficient pruning strategy, SyncScan, and an efficient representative pattern checking scheme, -dominate sequential pattern checking. The CRGSgrow is a two-step approach: in the first step, we obtain all closed repetitive sequential patterns as the candidate set of representative repetitive sequential patterns, and at the same time get the most of representative repetitive sequential patterns; in the second step, we only spend a little time in finding the remaining the representative patterns from the candidate set. An empirical study with both real and synthetic data sets clearly shows that the CRGSgrow has good performance.

## 1 Introduction

Sequential pattern mining has been a central data mining research topic in broad applications, including frequent sequential patterns-based classification, analysis of web log, analysis of frequent sequential patterns in DNA and protein sequence, API specification mining and API usage mining from open source repositories, and so on. So far many efficient sequential mining algorithms have been proposed for solving various of real problems, such as the general sequential pattern mining[3, 17, 27], frequent episode mining[16], closed sequential pattern mining[22, 25], maximal sequential pattern mining[14], top-k closed sequential pattern mining[21], long sequential pattern mining in noisy environment[26], constraint-based sequential pattern mining[18], frequent partial order mining[19], periodic pattern mining[28], etc..

**Table 1: An Example of Sequence Database**

| Sequence_id | $e_1$ | $e_2$ | $e_3$ | $e_4$ | $e_5$ | $e_6$ | $e_7$ | $e_8$ |
|---|---|---|---|---|---|---|---|---|
| $S_1$ | A | B | B | C | B | A | C | B |
| $S_2$ | B | A | C | B | A | C | B | C |

In recent years, some studies have focused on a novel problem of sequential pattern mining, mining repetitive gapped sequential patterns[7]. By gapped sequential patterns, it means a sequential pattern, which appears in a sequence in a sequence database, possibly with gaps between two successive events. In addition, for brevity, we use the term sequential pattern instead of gapped sequential patterns in this paper. Because traditional sequence mining[3, 17, 27] ignores the possibility of repetitive occurrences of sequential patterns in a sequence, it is a very important problem that discovers frequent repetitive sequential patterns by capturing not only repetitive occurrences of sequential patterns in different sequences but also repetitive occurrences within each sequence[7]. Currently a few approaches have been proposed to solve how to mine repetitive sequential patterns, but they cannot avoid an explosive number of output frequent repetitive sequential patterns for Apriori property, especially when the support threshold is low. Hence, it is difficult to understand the result set of frequent repetitive sequential patterns.

To solve the above challenge, it is natural to decrease the size of output result set, and two solutions [7, 14] have been proposed: *mining maximal repetitive sequential patterns*[1] and *mining closed repetitive sequential patterns*[2]. Mining maximal repetitive sequential patterns only focuses on the structure information of repetitive sequential pattern and fails to keep the information of the support. The mining closed repetitive sequential patterns emphasizes to have the same support of sub-sequence and super-sequence exactly (see the definitions in section2.1), resulting in the number of closed repetitive sequential patterns still too large to be used easily. Especially, mining closed repetitive sequential patterns will possibly generate more closed sequential patterns than traditional mining closed sequential patterns, since that the former studies not only the repetition in different sequences but also those repetitions within each sequence. The following example will further explain.

**Example 1:** Table 1 shows a sequence database $SeqDB = \{S_1, S_2\}$. If we compute the value of the support based on traditional sequence mining, the support of sequential pattern *AB* is 2 since *AB* occurs in each sequence in the database. In addition, the support of sequential pattern *ABB* is 2, and *AB* is a sequential pattern of *ABB*. Therefore, *ABB* is a closed sequential pattern in the database, and *AB* is not a closed sequential pattern. However, if we consider the same problem under the condition of the repetitive sequence mining, the non-overlapping occurrence of sequential pattern *AB* is 4 ($<e_1,e_2>$ and $<e_6,e_8>$ in $S_1$, $<e_2,e_4>$ and $<e_5,e_7>$ in $S_2$), and non-overlapping occurrence of sequential pattern *ABB* is 2 (only $<e_1,e_2,e_3>$ in $S_1$, $<e_2,e_4,e_7>$ in $S_2$), so both *AB* and *ABB* are closed repetitive sequential patterns.

---

[1] A frequent repetitive sequential pattern RP is maximal if there exist no a frequent super-sequence of RP.
[2] A frequent repetitive sequential pattern RP is closed if there exist no the super-sequence with the same support of RP.

**Table 2: An Example of Repetitive Sequential Patterns**

| Pattern_ID | Repetitive Sequential Patterns | Support |
|---|---|---|
| $RP_1$ | A, B, C | 51374 |
| $RP_2$ | A, B, C, D | 50961 |
| $RP_3$ | A, B, C, D, A | 26839 |
| $RP_4$ | A, B, C, D, A, B | 26747 |
| $RP_5$ | B, A, B, C, D, A, B | 26742 |

According to the above example, it is clear that mining closed repetitive sequential pattern is worse than the traditional mining closed sequential patterns in reducing the number of all frequent sequential patterns. Thus, apart from the two extremes of maximal repetitive sequential pattern and closed repetitive sequential pattern, we should find a solution to compress the result set of repetitive sequential patterns with a smaller number of representative repetitive sequential patterns. We give a motivating example as follows.

**Example 2:** Table 2 shows a subset of all frequent repetitive sequential patterns on a real dataset from a workflow system, where A, B, C, D are the names of distinct events. The five repetitive sequential patterns are all closed repetitive sequential patterns, since $RP_1 \subset RP_2 \subset RP_3 \subset RP_4 \subset RP_5$ and they have different supports between each other. The only $RP_5$ is maximal repetitive sequential pattern, since $RP_5$ contains the other patterns. And then, it is obvious that mining closed repetitive sequential patterns can not reduce the scale of the subset, and mining maximal repetitive sequential patterns can not express complete information of subset. However, it is clear that $RP_1$ and $RP_2$ are very similar with respect to structures and supports. Thus, $RP_1$ can be represented by $RP_2$. Similarly, $RP_3$, $RP_4$ and $RP_5$ have the same circumstance, so $RP_5$ also can represent $RP_3$ and $RP_4$. Therefore, $RP_2$ and $RP_5$ can represent the subset of the repetitive sequential patterns with carrying sufficient information and decreasing the size of the subset.

The above Example2 show that we can use a few representative repetitive sequential patterns to compress all frequent repetitive sequential patterns. Recently the problem of compressing frequent itemset has been studied [1, 23, 24], and two algorithms RPglobal and RPlocal have been presented. So, Can we compress repetitive sequential patterns by extension of the approach of compressing frequent itemset? Unfortunately, the answer cannot be so optimistic owing to the two following reasons. Firstly, there is not the one-step algorithm like RPlocal in mining repetitive sequential patterns with very small probability, since RPlocal is a greedy compressing frequent itemset algorithm based on the local boundary, which is not able to solve event order in a sequence. Secondly, although RPglobal algorithm may be applied to the problem of compressing repetitive sequential patterns, RPglobal consume high computational costs. Since RPglobal includes three steps: mining all frequent patterns; obtaining all cover information; calculating the minimal set-covering to find representative patterns, which will spend much time to collect complete covering information. *Hence, we should design an efficient and effective*

*algorithm to compress repetitive sequential patterns, whose efficiency is close to a one-step algorithm and effectiveness is equivalent to RPglobal.*

In this paper, we propose and study the problem of compressing repetitive gapped sequential patterns. Inspired by the ideas of summarizing frequent itemsets[23, 24], we firstly cluster all repetitive sequential patterns into a small number of groups whose members have similar structure and support, and then select a representative repetitive sequential pattern from each group. To obtain the high-quality compression, we propose a novel distance to measure the quality which shows the similarity between repetitive sequential patterns. Then, according to the distance threshold given by users, we define $\delta$-sequence cover in order to choose representative repetitive sequential patterns. Finally, the problem compressing repetitive gapped sequential patterns is equivalent to minimize the number of representative repetitive sequential pattern, which is formally reduced to minimal set-covering problem that is a well-known NP-Hard problem [23]. Since there is no polynomial time algorithm for the problem, we develop an algorithm, CRGSgrow, including an efficient pruning strategy, *SyncScan*, and an efficient representative pattern checking scheme, *$\delta$-dominate sequential pattern checking*. The CRGSgrow is a two-step approach: in the first step, we obtain all closed repetitive sequential patterns as the candidate set of representative repetitive sequential patterns, and at the same time get the most of representative repetitive sequential patterns; in the second step, we only spend a little time in finding the remaining the representative patterns from the candidate set. Empirical results prove that the running time of CRGSgrow is close to that of algorithm of mining closed repetitive sequential patterns, such as CloGSgrow[7], and much less than that of RPglobal.

The rest of the paper is organized as follows. In Section 2, we present some preliminaries about repetitive sequential patterns. The problem formulation will be introduced in Section 3. Section 4 focuses on the CRGSgrow algorithm, mainly introducing the SyncScan pruning strategy and the $\delta$-dominate sequential pattern checking scheme. We discuss our experimental results in Section 5, present the related work in Section 6 and give our conclusion in Section 7, respectively.

## 2 Preliminaries

To simplify our discussion, let us first introduce some preliminaries for repetitive gapped sequential patterns mining.

Let $I = \{i_1, i_2, ..., i_m\}$ be a set of distinct events. A sequence S is an ordered list of events, and denote S as $\langle e_1, e_2, ..., e_{end} \rangle$, where $e_i$ is an event, namely, $e_i \in I$ ($1 \leq i \leq end$). For brevity, a sequence of $S = \langle e_1, e_2, ..., e_{end} \rangle$ is also written as $S = e_1 e_2 ... e_{end}$. We refer to the $i^{th}$ event $e_i$ in the sequence S as S[i]. A sequence S is also a sequential pattern. The number of events in a sequence is called the length of the sequence and a sequence of length l is also called an l-sequence. A sequence database is a set of sequences, denoted by $SeqDB = \langle S_1, S_2, ..., S_N \rangle$.

A sequence $S_1 = e_1 e_2 ..., e_m$ is a **subsequence** of a sequence $S_2 = e'_1 e'_2 ... e'_n$, denoted by $S_1 \subseteq S_2$, if $m \leq n$ and there exist integers $1 \leq l_1 < l_2 < ... < l_m \leq n$ such that

$e_1 = e'_{l_1}, e_2 = e'_{l_2}, ..., e_m = e'_{l_m}$. We also call that $S_2$ is ***supersequence*** of $S_1$. The above a sequence of integers $\langle l_1, l_2...l_m \rangle$ is called a ***landmark*** of $S_1$ in $S_2$. For two patterns $P_1$ and $P_2$, if $P_1$ is a sequential patterns of $P_2$, we call that $P_1$ is a ***sub-pattern*** of $P_2$, and $P_2$ is a ***super-pattern*** of $P_1$.

In a sequence database $SeqDB = \langle S_1, S_2, ..., S_N \rangle$, there exists a pattern $P = e_1 e_2 ... e_m$ and its landmark is $\langle l_1, l_2...l_m \rangle$ in $S_i \in SeqDB$, the pair $(i, \langle l_1, l_2...l_m \rangle)$ is said to be an ***instance of P*** in $SeqDB$, and in particular, an ***instance of P*** in sequence $S_i$. Then, $S_i(P)$ is used to denoted ***the set of instance of P*** in $S_i$, and $SeqDB(P)$ is used to denote ***the set of instance of P*** in $SeqDB$. In addition, for a set of instances $I \subseteq SeqDB(P)$, we use $I_i$ to denote the subset of I containing the instances in $S_i$, and $I_i = I \cap S_i(P) = \{(i, \langle l_1^{(k)}, ..., l_m^{(k)} \rangle), 1 \leq k \leq n_i\}$. $n_i$ is the number of instances of P in the i$^{th}$ sequence. We will further explain the above concepts by the following example.

**Example 3:** Recall Table 1 to show a sequence database $SeqDB = \{S_1, S_2\}$. There are three landmarks of pattern AC in $S_1$ and five landmarks of pattern AC in $S_2$. Hence, there exist three instances of pattern AC in $S_1$: $S_1(AC) = \{(1, \langle 1, 4 \rangle), (1, \langle 1, 7 \rangle), (1, \langle 6, 7 \rangle)\}$ and five instances of pattern AC in $S_2$: $S_2(AC) = \{(2, \langle 2, 3 \rangle), (2, \langle 2, 6 \rangle), (2, \langle 2, 8 \rangle), (2, \langle 5, 6 \rangle), (2, \langle 5, 8 \rangle)\}$, and the set of instance of pattern AC in $SeqDB$: $SeqDB(AC) = S_1(AC) \cup S_2(AC)$.

Given two instances of pattern $P = e_1 e_2 ... e_m$ in $SeqDB = \{S_1, ..., S_N\}$, we said two instances, $(i_1, \langle l_1, l_2...l_m \rangle)$ and $(i_2, \langle l'_1, l'_2...l'_m \rangle)$, are ***overlapping instances*** if (i) $i_1 = i_2$ and (ii) $\exists 1 \leq j \leq m : l_j = l'_j$. This is equivalent that two instances, $(i_1, \langle l_1, l_2...l_m \rangle)$ and $(i_2, \langle l'_1, l'_2...l'_m \rangle)$, are ***non-overlapping instances*** if (i') $i_1 \neq i_2$ or (ii') $\forall 1 \leq j \leq m : l_j \neq l'_j$. According to the definition of the overlapping instances, we can define that a set of instances of pattern P in $SeqDB$ $I$ is ***non-overlapping instance set*** if any two instances in $I$ are non-overlapping instances. In addition, $(i_1, <l_1, l_2...l_m>)$ is believed that comes before $(i_2, <l'_1, l'_2...l'_m>)$ in ***the right-shift order*** if $(i_1 < i_2) \vee (i_1 = i_2 \wedge l_m < l'_m)$

Under consideration of repetitive sequential patterns pattern within one sequence or among multiple sequences, the ***repetitive support*** of a pattern P in $SeqDB$ is defined to be sup(p) = max $\{|I| | I \subseteq SeqDB(P)$ is non-overlapping instance set$\}$, and the non-overlapping instance set $I$ with $|I| = \sup(P)$ is defined a ***support set*** of $P$ in $SeqDB$. We will further explain above concepts about the overlapping of two instances, repetitive support and support set by the following example.

**Example 4:** Recall Example 3 and Table 1, two instances $(i_1, \langle l_1, l_2 \rangle) = (1, \langle 1, 4 \rangle)$ and $(i_2, \langle l_1', l_2' \rangle) = (1, \langle 1, 7 \rangle)$ are overlapping in the set of pattern AC in *SeqDB*, because $i_1 = i_2$ and $l_1 = l_1'$. On the contrary, two instances $(i_1, \langle l_1, l_2 \rangle) = (1, \langle 1, 4 \rangle)$ and $(i_2, \langle l_1', l_2' \rangle) = (1, \langle 6, 7 \rangle)$ are non-overlapping in the set of pattern AC in *SeqDB*, because $l_1 \neq l_1'$ and $l_2 \neq l_2'$. Accordingly, $I^{AC} = \{(1, \langle 1, 4 \rangle), (1, \langle 6, 7 \rangle), (2, \langle 2, 3 \rangle), (2, \langle 5, 6 \rangle)\}$ and $I'^{AC} = \{(1, \langle 1, 7 \rangle), (2, \langle 2, 6 \rangle), (2, \langle 5, 8 \rangle)\}$ are non-overlapping instance sets of pattern AC, so non-overlapping instance set is not unique. Based on the definition of repetitive support, $|I^{AC}| = 4$ is the maximum size of all non-overlapping instance sets, so $I^{AC}$ is the support set.

Based on the definition of repetitive support, we define that a pattern P is *frequent* if $\sup(P) \geq \min\_\sup$, where *min_sup* is specified by users. To sum up, the task of ***mining (closed) repetitive gapped sequential patterns*** is to discover all the frequent (closed) patterns in condition of SeqDB and min_sup.

## 3 Problem Formulation

In this section, we formally define the problem of compressing repetitive gapped sequential patterns. Firstly, we define a new distance measure based on Jaccard distance of repetitive gapped sequential patterns. Secondly, we put forward some concepts, such as $\delta$ - sequence cover and $\delta$ - dominate sequential pattern. Finally, we show the problem of compressing repetitive gapped sequential patterns is equivalent to minimal set-covering problem which is a well-known NP-Hard problem.

If a repetitive sequential pattern $P_1$ can properly represent another repetitive sequential pattern $P_2$, $P_1$ must be similar to $P_2$. To measure the similarity between two repetitive gapped sequential patterns, we need a reasonable measurement criterion. Ref. [23] proposed a method based on Jaccard distance to measure the similarity between two itemsets, so it is natural to use the distance to measure the similarity between two repetitive gapped sequential patterns. Unfortunately, the measurement criterion can not been adopted directly for the characteristic of repetition of sequential pattern within each sequence. Hence, we propose a novel Jaccard distance for measuring the similarity between repetitive gapped sequential patterns.

**Definition 1 (Distance between two repetitive gapped sequential patterns)** Let $P_1$ and $P_2$ be two repetitive gapped sequential patterns. The distance of $P_1$ and $P_2$ is defined as:

$$D(P_1, P_2) = 1 - \frac{\text{min-ins}\{S(P_1) \bigcap S(P_2)\}}{\text{max-ins}\{S(P_1) \bigcup S(P_2)\}}$$

Where S(P) is the set of sequences in the given sequence database which contains the sequential pattern P, min-ins$\{S(P_1) \bigcap S(P_2)\}$ is the least support between $P_1$ and $P_2$ in every sequence including them, and max-ins$\{S(P_1) \bigcap S(P_2)\}$ is the most support between $P_1$ and $P_2$ in every sequence including them.

**Example 5:** Recall Table 1, let $P_1$=AB and $P_2$=ABC, $S(P_1) = S(P_2) = \{S_1, S_2\}$, where $S_i$ is a sequence in the sequence database. In $S_1$, the non-overlapping instance set of $P_1$ is $I_1^{AB} = \{(1,\langle 1,3\rangle),(1,\langle 6,8\rangle)\}$, $|I_1^{AB}|=2$, and the non-overlapping instance set of $P_2$ is $I_1^{ABC} = \{(1,\langle 1,2,4\rangle)\}$, $|I_1^{ABC}|=1$. The lesser number of instances is 1 in $S_1$. For the same reason, $I_2^{AB} = \{(2,\langle 2,4\rangle),(2,\langle 5,7\rangle)\}$, $|I_2^{AB}|=2$, $I_2^{ABC} = \{(2,\langle 2,4,6\rangle),(2,\langle 2,7,8\rangle)\}$, $|I_2^{ABC}|=2$, the lesser number of instances is 2 in $S_2$. Hence, min-ins$\{S(P_1)\bigcap S(P_2)\}$ =1+2=3. According to the definition of max-ins$\{S(P_1)\bigcap S(P_2)\}$, max-ins$\{S(P_1)\bigcap S(P_2)\}$ =2+2=4, and then $D(P_1,P_2) = 1-3/4 = 1/4$.

**Theorem 1.** The distance between two repetitive gapped sequential patterns D is a distance metric.

**Sketch of Proof.** See Appendix. □

Intuitively, in a set of sequential patterns Seq= $\{sp_1, sp_2,...,sp_n\}$, if a sequential pattern RP can properly represent a set of sequential patterns, then $\forall sp_i \in Seq$, $sp_i \subseteq RP$ and $D(sp_i, RP) \leq \varepsilon$ ($1 \leq i \leq n$), where $\varepsilon$ is a very small positive real number. Inspired by the concept of $\delta$-cover in [23], we use similar definition of $\delta$-sequence cover to formulate the above intuition. Note $\delta$ is a threshold of distance between two repetitive sequential patterns specified by users and $\delta \in [0,1]$.

**Definition 2 ($\delta$-sequence cover)** A repetitive sequential pattern $P$ is $\delta$-sequence covered by another repetitive sequential pattern $RP$ if $P \subseteq RP$ and $D(P,RP) \leq \delta$ ($\delta \in [0,1]$).

According to the definition of $\delta$-sequence cover, a repetitive sequential pattern P must be a sequential patterns of a representative sequential pattern if the representative sequential pattern can $\delta$-sequence cover P. Therefore, we can simplify the above definition of distance between two repetitive gapped sequential patterns: $D(P,RP) = 1 - \dfrac{\text{min-ins}\{S(P)\bigcap S(RP)\}}{\text{max-ins}\{S(P)\bigcup S(RP)\}} = 1 - \dfrac{\sup(RP)}{\sup(P)}$.

Since checking $\delta$-sequence cover between any two sequential patterns will spend much time on computing, we will introduce some novel properties about compressing repetitive sequential pattern to speed up the checking $\delta$-sequence cover.

**Definition 3(min sequence cover)** Given a set of repetitive sequential patterns S, min sequence cover of SP (MSC(SP) in short), a repetitive sequential pattern in S, is define as follows:

$$\text{MSC(SP)} \begin{cases} \min\{D(SP,SP_i)|\forall SP_i \in S, SP \subseteq SP_i\} & \exists SP_i \in S \\ +\infty & \forall SP_i \notin S \end{cases}$$

**Definition 4($\delta$-dominate sequential pattern)** Given a set of repetitive sequential patterns S, SP is a repetitive sequential patterns in S. SP is a $\delta$-dominate sequential pattern in S, if MSC(SP)>$\delta$. Equivalently, SP can not be $\delta$-sequence cover by any repetitive sequential patterns in S, if MSC(SP)>$\delta$.

We will further explain two above definitions in the following example. For example, given a set of repetitive sequential patterns S={(AB:100), (AC:80), (AD:60), (ABC:60), (ABD:50), (ABCD:20), (BABCD:5)} and $\delta$ =0.3. There are some super-sequence of AB, such as ABC, ABCD and BABCD. MSC(AB)= $D(AB, ABC)$ $=1-60/100=1-0.6=0.4<0.3$, then pattern AB is not a $\delta$-dominate sequential pattern. On the other hand, there exists only super-sequence of ABCD in S, BABCD. MSC(ABCD)= $D(ABCD, BABCD)=1-5/20=1-0.25=0.75>0.3$, then pattern ABCD is a 0.3-dominate sequential pattern, namely, there are not any pattern can 0.3-sequence cover pattern ABCD in S.

**Lemma 1.** Given a set of repetitive sequential patterns S, all closed repetitive sequential patterns in S are *0-dominate sequential pattern*, and all maximal repetitive sequential patterns in S are *1-dominate sequential pattern*.

Proof: Directly from the definition closed repetitive sequential pattern and maximal repetitive sequential pattern. □

**Theorem 2** Given a set of repetitive sequential patterns S, and any set of representative repetitive sequential patterns, RS, which can $\delta$ sequence cover S. Then, RS must contain all $\delta$-dominate sequential patterns in S.

**Proof:** If DP is any $\delta$-dominate sequential pattern in S, based on the definition of $\delta$-dominate sequential pattern, the MSC(DP) in S must larger than $\delta$. Thus, DP can not be $\delta$ sequence covered by any repetitive sequential pattern in S. So, DP must be a representative sequential pattern in any set of representative sequential patterns RS that can $\delta$ sequence cover S. □

**Definition 5(Repetitive Gapped Sequential patterns Compression)** Given a sequence database SeqDB, a minimum support min_sup and distance threshold $\delta$, the compressing repetitive gapped sequential patterns is to find a set of representative repetitive gapped sequential pattern RRGS, such that for each frequent repetitive gapped sequential pattern P (w.r.t min_sup), there exits a representative repetitive gapped sequential pattern $RP \in RRGS$ (w.r.t min_sup) which $\delta$-sequence cover P, and the |RRGS|, which the size of set of representative repetitive gapped sequential pattern RRGS, is minimized.

**Theorem 3** The problem of compressing repetitive gapped sequential patterns is NP-Hard.

Following similar concepts of frequent itemset compression, we also formulate the problem of compressing repetitive gapped sequential patterns to the problem of minimal set-covering, a well-known NP-Hard problem. Due to the space limit, we omit the proof in this paper and refer reader to [23] for the proof. □

## 4 Efficient Compressing Repetitive Gapped Sequential Patterns Algorithm

In this section, given a sequence database SeqDB, a minimum support min_sup, and a distance measure threshold $\delta$, we elaborate an algorithm CRGSgrow for **c**ompressing **r**epetitive **g**apped **s**equential patterns. The CRGSgrow is a two-step approach: in the first step, we obtain all closed repetitive sequential patterns as the candidate set of representative repetitive sequential patterns, and at the same time get all $\delta$-dominate sequential patterns; in the second step, we only spend a little time in finding the remaining the representative patterns from the candidate set. We start with introducing the design and implementation of the algorithm CRGSgrow in subsection 4.1. Then, we analyze the time complexity of all our algorithms in subsection4.2.

**4.1 CRGSgrow: Design and Implementation**

In this section, we firstly introduce the algorithm, CRGSgrow, for compressing repetitive gapped sequential patterns.

**Algorithm 1:** CRGSgrow

**Input**: sequence database **SeqDB**={ $S_1, S_2, ..., S_n$ }; threshold **min_sup**; a distance threshold $\delta$

**Output:** A set of representative repetitive sequential patterns

**1:** $E \leftarrow$ all frequent 1-sequential patterns in SeqDB; Cover $\leftarrow \varnothing$; Covered $\leftarrow \varnothing$;
**2: for each** $e \in E$ **do**
**3:**   **if** $e$ is visited **then** continue;
**4:**   $P \leftarrow e$; $I \leftarrow \{(i, <l>) \mid \text{for some } i, S_i[l] = e\}$;
**5:**   $P' \leftarrow \varnothing$; $I' \leftarrow \varnothing$
**6:**   SyncScan (SeqDB, $P$, $I$, $P'$, $I'$, Cover, Covered);
**7:** Cover $\leftarrow$ Compress (Cover, Covered);
**8: return** Cover;

In algorithm1, the CRGSgrow traverses the pattern space in a depth-fisrt way. We first get all the 1-sequential patterns ordered by supports. All 1-sequential patterns with their support sets are found (line 4), and then SyncScan(SeqDB, P, I, $P'$, $I'$ Cover, Covered) is called (line 6) to find all closed sequential patterns with P as their prefixes. In the process of SyncScan, the Cover and Covered sets will be also updated continuously. Especially, all $\delta$-dominate sequential patterns will be obtained in this process. At last, Compress(Cover, Covered) will finish all the compression work, and will be explained in algorithm 2.

**Algorithm 2:** Compress

**Input**: $\delta$-dominate sequential patterns set **Cover**; other closed frequent sequential patterns set **Covered**

**Output:** A set of compressed repetitive sequential patterns set

**1: for each** *RP* in *Cover* **do**
**2:**   **for each** *P* in *Covered* **do**
**3:**     **if** *RP* can $\delta$ sequence cover *P* **then**
**4:**       Put *P* into Set *T*;
**5:**   $\bar{T} \leftarrow$ *Cover - T*
**6: for each** *SP* in $\bar{T}$ **do**
**7:**   **for each** *CR* in *Covered-Cover* **do**
**8:**     **if** *CR* can $\delta$ sequence cover *SP* **then**
**9:**       Put *SP* into Set(*CR*);
**10: While** $\bar{T} \neq \varnothing$ **do**
**11:**   select a sequential pattern *CR* which can maximize | Set(*CR*)|
**12:**   **for each** $SP \in \text{Set}(CR)$ **do**
**13:**     Remove *SP* from $\bar{T}$ and other Set(*CR'*) (*CR'* $\in$ *Covered-Cover*)
**14: return** Cover

In algorithm2, we firstly find all the patterns in the set Covered which can be $\delta$-sequence covered by the $\delta$-dominate sequential patterns in the set Cover (line 1~3). These patterns will be put into set T (line 4). Then all the other patterns in the set Covered except patterns in set T will be put into set $\bar{T}$. For these patterns, we can deal with them by the set-covering greedy algorithm which iteratively discovers every representative repetitive sequential pattern [23].

In algorithm 2, the inputs are two sets, $\delta$-dominate sequential patterns set and other set including all other frequent closed patterns except $\delta$-dominate sequential patterns. In Definition 4, we know that $\delta$-dominate sequential patterns are frequent closed sequential patterns which can not be $\delta$-sequence covered by its super-pattern. Thus, all the $\delta$-dominate sequential patterns must be the representative patterns. Hence, we will introduce the SyncScan algorithm in the algorithm 3 as follows.

Now one problem remains: how to obtain $\delta$-dominate sequential patterns and mining closed repetitive sequential patterns efficiently. In this paper, we will propose an algorithm called SyncScan to solve this problem. The process of searching $\delta$-dominate sequential patterns will be conducted together with the process of closed sequential patterns mining in this algorithm. Meanwhile, a reasonable pruning strategy will be applied to this process to improve the efficiency of the algorithm. Hence, we will introduce the SyncScan algorithm in algorithm 3.

In algorithm 3, the algorithm is a depth-first search algorithm of the pattern space starting from P (or $P'$), to find all frequent patterns with P (or $P'$) as prefixes and put them into the proper set. At first we will do the closed check of the current pattern using the subroutine named Check(P). Then we will judge that if the length of the current pattern P is 2, and all the sequences including the current pattern P contains all the sequences including the second event ev of the current pattern. If the condition holds, we will begin to search the patterns prefixed with ev simultaneously. Otherwise we only search the patterns prefixed with the current pattern. In addition, in each iteration of line 6~12, support set I (or $I'$) of pattern P∘e (or $P'$∘e) is obtained. This will be recursively called in line 9 or line 12. The process of obtaining $\delta$-dominate sequential patterns will be conducted in subroutine Check.

In this algorithm, we will first check if the current pattern can be pruned using LBCheck[7]. Then if the current pattern is closed, we will do the $\delta$-dominate checking with the method, DomCCheck(P). If the current pattern is $\delta$-dominate sequential pattern, it will be put into the set Cover. Otherwise, the closed pattern will be put into the set Covered.

To sum up, we will take an example to further explain the pruning strategy above. Since the biggest difference between repetitive sequence mining and traditional sequence mining is that repetitive sequence mining needs to capture not only repetitive occurrences of sequential patterns in different sequences but also repetitive occurrences within each sequence. Thus, computing the support of repetitive sequential pattern will consume more time. If we may prune a part of search space, the process of compressing repetitive sequential patterns will speed up rapidly.

**Algorithm 3:** SyncScan
**Input**: sequence database **SeqDB**={ $S_1, S_2, ..., S_n$ }; threshold **min_sup**; Pattern P=$e_1, e_2, ... e_{j-1}$; $P' = e_1, e_2, ... e_{j-1}$ or $\emptyset$; leftmost support set $I$ of $P$ in SeqDB; semi-left support set $I'$ of pattern $P'$ in SeqDB; $\delta$-dominate sequential patterns set, Cover; a set of other closed repetitive sequential patterns, Covered
**Output:** Cove; Covered
1: Check*(P, I, Cover, Covered)*;
2: **if** $P' \neq \emptyset$ **then**
3:     $P' \leftarrow \{P - e_1\} \cup P'$; Check*( $P'$, $I'$, Cover, Covered)*;
4: $ev \leftarrow$ the second event of P;
5: **for each** $e \in \alpha$ **do**
6:    **if** length of $P=2$ and all ev occurs in the same sequence with $P$ **then**
7:        $P' \leftarrow ev$; $I' \leftarrow \{(i, <l>) |$ for some i, $S_i[l] = ev$ except instances in P};
8:        obtaining two non-overlapping instance sets of $I^+$ and $I^{+'}$ with e;
9:        SyncScan(*SeqDB, P∘e, $I^+$, $P' \circ e$, $I^{+'}$*);
10:   **else** $P' \leftarrow \emptyset$, $I' \leftarrow \emptyset$;
11:       obtaining the non-overlapping instance sets of $I^+$ with e;
12:       SyncScan(*SeqDB, P∘e, $I^+$, $P' \circ e$, $I^{+'}$*);
**Subroutine Check( $P$ )**
**Input:** sequence database **SeqDB**; Pattern P $\delta$-dominate sequential patterns set, Cover; a set of other closed repetitive sequential patterns, Covered
**Output:** Cover, Covered
13: **if** |I|$\geq$ min_sup && LBCheckprune(P) and DomCCheck(P) $\neq$ nclosed **then**
14:    **if** DomCCheck( $P$ )= $\delta$-dominate **then**
15:        Cover $\leftarrow$ Cover $\cup P$;
16:    **else** Covered $\leftarrow$ Covered $\cup P$;

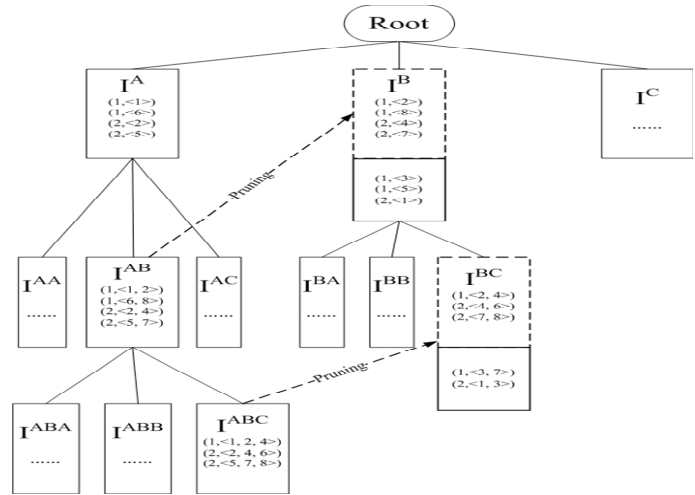

**Fig. 1 SyncScan Pruning Strategy**

**Example 6:** Recall Table 1 shows a sequence database $SeqDB = \{S_1, S_2\}$. We will compute sup(ABC) and sup(BC) simultaneously in the way illustrated in Figure 1. The complete search space of computing support of each repetitive sequential pattern forms the lexicographic sequence tree shown in Figure1. The intuition of our pruning strategy is that subsequence of searched sequential patterns don't need to repeat the same searching process. Thus, we can compute supports of both ABC and BC in the same process, since BC is a subsequence of ABC. We will explain each step as follows:

1) Find a support set $I^A$ of A. $I^A$ is the set of all instances of A and is shown the rectangle labeled $I^A$.

2) Find a support set $I^{AB}$ of 'AB'. We can extend each instance in $I^A$, adding the next 'B' after current event 'A' in the right-shift order[7]. Meanwhile, we also start to compute the support of the patterns with the prefix of 'B' which are not contained in instances of the pattern 'AB' in the sequence database. The process is based on the right-shift order. Those instances of patterns prefixed with 'B' included in the instances of patterns prefixed with 'AB' have been searched in the process of extending the pattern 'AB'. Thus, these patterns will not be searched again in the process of extending the pattern 'B'. These patterns will be pruned and marked as the dotted rectangle of $I^B$ in Figure 1.

3) Find a support set $I^{ABC}$ of ABC. Similar to step 2, for there is no 'C' to be extended for (1, <6, 8>) in $S_1$, we stop extending (1, <6, 8>). For (1, <1, 2>) in $S_1$, it can be extended to (1, <1, 2, 4>). Then we will continue searching in $S_2$. Note we also start to compute sup(BC) when computing sup(ABC). To avoid repeating search of the patterns prefixed with 'ABC', we only can search the remaining patterns prefixed with 'BC', and finally accumulate them. In the end, we finish computing sup(ABC) =3 and sup(BC)=5 in the same process.

### 4.2 Time Complexity Analysis

In this subsection, we analyze the time complexity of our mining algorithms CRGSgrow. We first analyze the time complexity of the compression algorithm, Compress, and then analyze the complexity of CRGSgrow.

**Lemma 2**(Time Complexity of Compress): Algorithm of Compress's time complexity is $|Cover| \cdot |Covered| + |\bar{T}| \cdot |Covered| + O(\sum_{CR \in Cover-Covered} Set(CR))$.

**Proof:** In the first step (line1~4), algorithm 2, Compress, needs to conduct at most $|Cover| \cdot |Coverd|$ sequential patterns testing. In the second step (line 6~9), the algorithm 2 needs to conduct at most $|\bar{T}| \cdot |Coverd|$ sequential patterns testing. The original RPglobal needs to conduct at most $|cover \cup covered|^2$ sequential patterns testing. Clearly, $|Cover| \cdot |Covered| + |\bar{T}| \cdot |Covered| < |cover \cup covered|^2$.

In the third step (line10~13), it is the set-covering greedy algorithm which iteratively selects a representative sequential pattern which covers as many closed frequent sequential patterns as possible. This step can be implemented in time complexity $O(\sum_{CR \in Cover-Covered} Set(CR))$ [23]. □

To analyze the time complexity of algorithm 1, CRGSgrow conveniently, we define F= Cover $\cup$ Covered as all the closed iterative sequential patterns and define N = $|E|$ as the number of distinct events. $O(T)$ represents the time complexity of Compress.

**Lemma 3** (Time complexity of CRGSgrow) The Algorithm of CRGSgrow's time complexity is $O(\sum_{P \in F} \sup(P) \cdot N \log L) + O(T)$.

**Proof:** In algorithm 3, SyncScan, there is the process of obtaining the support set (line 8, 11) in algorithm 3. In [7], this process uses $O(\sup(P) \cdot \log L)$ time. But our method will be more efficient because of the pruning strategy applied. For example, in algorithm 2, if $P'$ is not $\varnothing$, pattern $P'$ will be extended with pattern P simultaneously. Because they can share many same instances in the process of pattern growth. This pattern growth process of the two patterns in [7] cost $(\sup(P) + \sup(P')) \cdot \log L$ time, but in our method, it will only cost $\max\{\sup(P), \sup(P')\} \cdot \log L$ time.

In algorithm 3, the time complexity of Check is determined by the number and structure of closed patterns. It has been studied in [7].

For each P $\in F$ and e $\in \alpha$, from the Apriori property and the algorithm 3, we know SyncScan is executed only for patterns in F.

According to the Pruning strategy in algorithm 3, In the best case, the time complexity is $O(\sum_{P \in F} \sup(P) \cdot \log L) + O(T)$. In the worst case, the time complexity is $O(\sum_{P \in F} \sup(P) \cdot \log L) + O(T) = O(\sum_{P \in F} \sup(P) \cdot E \log L) + O(T)$. So the total time is $O(\sum_{P \in F} \sup(P) \cdot E \log L) + O(T)$. □

## 5. Empirical Results

### 5.1 Test Environment and Datasets

In this section, we report a systematic performance study on both real data sets and synthetic data sets. All of our experiments were performed on a Lenovo ThinkPad T60 with Intel 4200 CPU, 1GB memory and Windows XP professional installed.

Algorithms were implemented in Microsoft Visual C++ V6.0. In the experiments, we compared CRGSgrow with a fast closed repetitive gapped sequential pattern mining algorithm, CloGSgrow, and another compressing frequent patterns algorithm, RPglobal, which were implemented for compressing repetitive gapped sequential patterns by us, using both real data sets and synthetic data sets.

The characteristics of the selected data sets are shown in Table 3.

**Table3 Real and Synthetic Data Sets Characteristics**

| Dataset | Type of Dataset | Number of Sequences | Number of Items | Average Sequence Length | Maximum Sequence Length |
|---|---|---|---|---|---|
| Gazelle | Real | 29369 | 1423 | 3 | 651 |
| TCAS | Real | 1578 | 75 | 36 | 70 |
| D5C20N10S20 | Synthetic | 5 | 10 | 20 | 20 |

The first data set, *Gazelle*, contains 29,369 web click-stream sequences from customers and 1423 distinct items, which has been a benchmark dataset used by past studies on sequential pattern mining. Although the dataset is sparse since the average sequence length is only, there are a large number of long sequences (the maximum length is 651), where a sequential pattern may repeat many times. More detailed information about this data set can be found in [9].

The second data set, TCAS dataset, is a set of software traces collected from Traffic alert and Collision Avoidance System. The dataset contains 1578 sequences and 75 distinct items. The average sequence length of the dataset is 36 and maximum sequence length is 70. More detailed information about this data set can be found in [11].

The third data set, D5C20N10S20, is a synthetic set generated by IBM sequence data generator [3]. The data generator requests a set of parameters, D, C, N and S, corresponding to the number of sequences, the average sequence length, the number of distinct items, and the maximum sequence length respectively, to produce the dataset.

We carry out our experiments to compare three algorithms, CloGSgrow, RPglobal, CRGSgrow, in the above three dataset mainly on the compression quality and running time. For each comparison, we vary the value of support threshold and fix $\delta = 0.2$ (it is a reasonably good compression quality).

In the experiments of compression quality, three algorithms, CloGSgrow, RPglobal, CRGSgrow, are compared with respect to the number of output repetitive gapped sequential patterns respectively. In addition, to verify the effectiveness of $\delta$-dominate sequential patterns we proposed in our work, we also make the experiments on the number of $\delta$-dominate sequential patterns which are showed as pink line in figure 2-4. Moreover, if a algorithm cannot finish within 60 minutes, we do not show the results, so we only give the partial results of RPgolbal because the running time of RPgolbal can not be tolerable when the minimum support threshold is low.

In the experiments of running time, we compare the running time of the three algorithms shown in figures 5-7. Especially, the runtime of RPglobal includes the time that generates all closed repetitive gapped sequential patterns by CloGSgrow and finds the set of representative repetitive gapped sequential patterns.

**5.2 Compression Quality**

As the figures 2-4 shown, we have the following observations: firstly, the number of representative repetitive gapped sequential patterns by CRGSgrow is a little more than the number of patterns generated by RPglobal, and the number of patterns outputted by CRGSgrow is about one-quarter of that closed repetitive gapped sequential patterns mined by CloGSgrow; secondly, in the algorithm of CRGSgrow, we can obtain the $\delta$-dominate sequential patterns which include the most of representative repetitive sequential patterns.

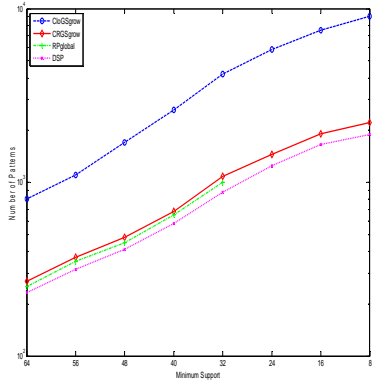

**Fig2. Num of Patterns of Gazelle**

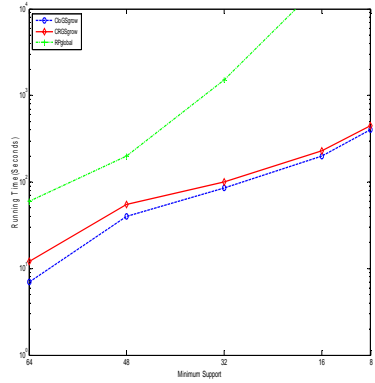

**Fig5. Running Time in Gazelle**

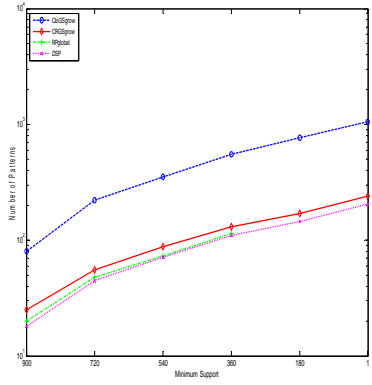

**Fig3. Num of Patterns of TCAS**

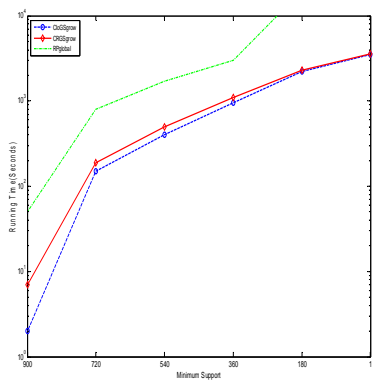

**Fig6. Running Time in TCAS**

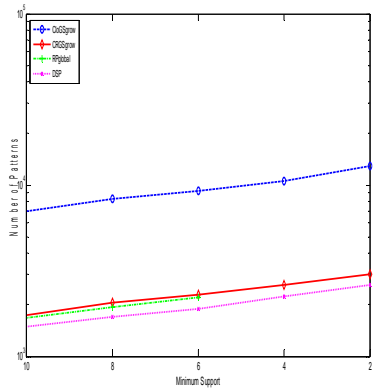

**Fig4. Num of Patterns of D5C20N10S20**

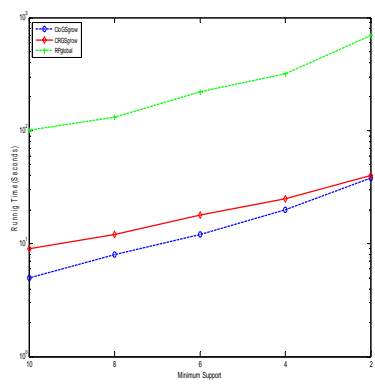

**Fig7. Running Time in D5C20N10S20**

### 5.3 Running Time

As the figures 5-7 shown, the running time of CRGSgrow is much less than the time of RPglobal, and is very close to the time of CloGSgrow. The observation explains SyncScan search space pruning strategy and $\delta$-dominate sequential patterns in the algorithm of CRGSgrow can save much time in finding the representative repetitive sequential patterns from the closed repetitive sequential patterns, to improve efficiency dramatically.

### 6. Related Work

To the best of our knowledge, this is the first study on compressing the repetitive gapped sequential patterns. In the following, we review some related work on mining sequential patterns, mining repetitive or gap-constrained patterns and compressing or summarizing frequent patterns.

There is a wealth of literature devoted to mining frequent patterns over sequence data. This problem was first introduced by Agrawal and Srikant in [3]. Since then, many sequential pattern mining algorithms have been proposed for performance improvement, such as SPADE by Zaki[27], PrefixSpan by Pei et al.[17] and SPAM by Ayres et al.[5]. Owing to the downward closure property of Apriori leads to an explosive number of frequent patterns, sequential pattern mining also confronts the severe challenge. Following similar concepts in frequent itemset mining [6, 16], the data mining community has proposed algorithms to mine closed sequential patterns[22, 25] and maximal sequential patterns[28] to remove the redundancy among sequential patterns. The concept of closed sequential pattern is proposed to keep all information about the sequential pattern in a lossless way, where both structure and support are fully preserved. In contrast, the approach of mining maximal sequential pattern only takes into those longest ones account, because all other sequential patterns must be contained by them.

Different from traditional sequential pattern mining, there are studies that focus on mining repetition of sequential patterns within sequences and mining sequential patterns satisfying gap-constrained recently [7, 8, 10, 11, 15, 20, 28]. Mining repetition of sequential patterns was first paid attention to by Mannila et al. [15], where a sequential patterns is called an episode. Mining episode focuses on finding all events occurring close to one another, and measuring closeness between two episodes with a fix-width window. Since then, Casas-Garriga et al. [8] change the measurement of closeness of episodes from fix-width window to a gap constraint. Later, Zhang et al. [28] introduced the concept of "gap requirement" in mining periodic patterns from single genome sequence, and both overlapping and non-overlapping periodic patterns all were discovered. Then, Li and Wang [10] represented an efficient algorithm, Gap-BIDE, can mine closed sequential patterns satisfying gap-requirement over multiple sequences. In recent, Lo et al. [11] and Ding et al. [7] proposed iterative pattern mining and closed repetitive gapped sequential patterns mining respectively, and there exists on the difference of support of instance about their works. The former captured occurrences based MSC/LSC semantics, and the latter captured occurrences based non-overlapping definition. In our work, we adopt the way of the latter to define support of a sequential patterns instance.

**Table 4 Related Works**

| Works | DataType | Repetitive patterns | Search Strategy | Removing Redundancy Approach and Constraint of Instances |
|---|---|---|---|---|
| Agrawal and Srikant | Multiple Sequences | Ignore | Breadth-first Search | All sequential patterns |
| Manilla et al. | Single Sequence | Consider | Breadth-first Search | All sequential patterns satisfying fixed-width windows or minimal windows |
| Wang et al. | Multiple Sequences | Ignore | Depth-first Search | Closed Sequential patterns |
| Zhang et al. | Single Sequence | Consider | Breadth-first Search | All sequential patterns satisfying gap-constrained |
| Lo et al. | Multiple Sequences | Consider | Depth-first Search | Closed Sequential patterns based MSC/LSC semantics |
| Li and Wang | Multiple Sequences | Consider | Depth-first Search | Closed sequential patterns satisfying gap-constrained |
| Ding et al. | Multiple Sequences | Consider | Depth-first Search | Closed non-overlapping sequential patterns |
| Dong et al. | Transaction Database | Ignore | Depth-first Search | Representative Itemset satisfying similar cluster |
| Our Work | Multiple Sequences | Consider | Depth-first Search | Non-Redundant and non-overlapping sequential patterns |

Although closed sequential pattern and maximal sequential pattern can partially alleviate this redundancy problem of sequential pattern mining, and then mining repetitive and gap- constrained sequential pattern also adopt this strategy to remove redundancies among sequential instances, there exist a large number of redundant patterns shown in Example 2. Since frequent itemset mining also encounter the above similar problem, some efficient algorithms [1, 23, 24] for compressing or approximating the collection of frequent itemsets to really eliminate redundancy of among itemsets have been proposed recently. Inspired by the ideas of compressing the collection of frequent itemsets, we consider whether maybe compressing the collection of the repetitive gapped sequential patterns. Unfortunately, the approach of compressing collection of frequent itemsets cannot be extended simply for the ordered feature of sequential pattern and repetitive property of repetitive gapped sequential patterns. Hence, we hope, apart from two extremes of maximal sequential pattern and closed sequential pattern, we can find a road in the middle, which differs from the maximal method in that it merge the special support of repetitive sequence in consideration, and also differs from the closed method because sub-sequence and super-sequence are not required to have exactly the same support so that they can be summarized.

In Table 4, some important features of our work are compared with other related works of different types.

## 7. Conclusion

This paper studies how to effectively and efficiently compress repetitive gapped sequential patterns from sequence database. To the best of our knowledge, the problem of compressing the repetitive gapped sequential patterns has not been well studied in existing work.

In this paper, we propose and study the problem of compressing repetitive gapped sequential patterns. Inspired by the ideas of compressing frequent itemsets, we firstly cluster all repetitive sequential patterns into a small number of groups whose members have similar structure and support, and then select a representative repetitive sequential pattern from each group. To obtain the high-quality compression, we propose a novel distance to measure the quality which shows the similarity between repetitive sequential patterns. Then, according to the distance threshold given by users, we define $\delta$-sequence cover in order to choose representative repetitive sequential patterns. Finally, the problem compressing repetitive gapped sequential patterns is equivalent to minimize the number of representative repetitive sequential pattern, which is formally reduced to minimal set-covering problem that is a well-known NP-Hard problem. Since there is no polynomial time algorithm for the problem, we develop an algorithm, CRGSgrow, including an efficient pruning strategy, *SyncScan*, and an efficient representative pattern checking scheme, *$\delta$-dominate sequential pattern checking*. Empirical results prove that the algorithm CRGSgrow can obtain a good compressing quality efficiently.

## References

1. FN. Afrati, A. Gionis, H. Mannila. Approximating a collection of frequent sets. In KDD 2004.
2. R. Agrawal and R. Srikant. Fast algorithms for mining association rules. In VLDD 1994.
3. R. Agrawal and R. Srikant. Mining sequential patterns. In ICDE 1995.
4. G. Ammons, R. Bodik, and J. R. Larus. Mining specification. In SIGPLAN POPL 2002.
5. J. Ayres, J. Flannick, J. Gehrke, and T. Yiu. Sequential pattern mining using a bitmap representation. In KDD 2002.
6. R. J. Bayardo. Efficiently mining long patterns from databases. In SIGMOD 1998.
7. B. Ding, D. Lo, J, Han, S.-C. Khoo. Efficient mining of closed repetitive gapped sequential patterns from a sequence database. In ICDE 2009.
8. G. Garriga. Discovering unbounded episodes in sequential data. In PKDD 2003.
9. R. Kohavi, C. Brodley, B. Frasca, L. Mason, and Z. Zheng KDD Cup 2000 Organizers' Report: Peeling the Onion. SIGKDD Explorations, vol. 2: 86-98, 2000.
10. C. Li and J. Wang. Efficiently mining closed sequential patterns with gap constraints. In SDM 2008.
11. D. Lo, S.-C. Khoo, and C. Liu. Efficient mining of iterative patterns for software specification discovery. In KDD 2007.
12. D. Lo and S.-C. Khoo. SMArTIC: Toward building an accurate, robust and scalable specification miner. In SIGSOFT FSE 2006.
13. D. Lo, S. Maoz, and S.-C. Khoo. Mining modal scenario-based specifications from execution traces of reactive systems. In ASE 2007.
14. C. Luo, S.M. Chung A scalable algorithm for mining maximal frequent sequences using a sample. Knowledge and Information System, 15:149 –179, 2008.

15. H. Mannila, H. Toivonen, and A. I. Verkamo. Discovery of frequent episodes in event sequences. Data Mining Knowledge. Discovery, vol. 1, no. 3: 259–289, 1997.
16. N. Pasquier, Bastide Y, Taouil R, Lakhal. Discovering frequent closed itemsets for association rules. In ICDT 1999.
17. J. Pei, J. Han, B. Mortazavi-Asl, H. Pinto, Q. Chen, U. Dayal, and M.-C. Hsu. Prefixspan: Mining sequential patterns efficiently by prefixprojected pattern growth. In ICDE 2001.
18. J. Pei, J. Han, and W. Wang Constraint-Based Sequential Pattern Mining in Large Databases In CIKM 2002.
19. J. Pei, J. Liu, H. Wang, K. Wang, P.S. Yu, and J. Wang. Efficiently Mining Frequent Closed Partial Orders. In ICDM 2005.
20. M. El-Ramly, E. Stroulia, and P. Sorenson. From run-time behavior to usage scenarios: an interaction-pattern mining approach In KDD 2002.
21. P. Tzvetkov, X. Yan, and J. Han. TSP: Mining Top-K Closed Sequential Patterns In ICDM 2003.
22. J. Wang and J. Han. BIDE: Efficient mining of frequent closed sequences. In ICDE 2004.
23. D. Xin, J. Han, X .Yan and H. Cheng. On compressing frequent patterns. KIS, 60: 5-29, 2007.
24. X .Yan, H. Cheng, J. Han and D. Xin. Summarizing Itemset Patterns: A Profile Based Approach. In KDD 2005.
25. X. Yan, J. Han, and R. Afhar. CloSpan: Mining closed sequential patterns in large datasets. In SDM 2003.
26. J. Yang, P.S. Yu, W. Wang, J. Han. Mining Long Sequential Patterns in a Noisy Environment. In SIGMOD 2002.
27. M. Zaki. SPADE: An Efficient Algorithm for Mining Frequent Sequences. Machine Learning, 42:31-60, Kluwer Academic Pulishers, 2001.
28. M. Zhang, B. Kao, D. Cheung, K. Yip. Mining Frequent periodic patterns with gap requirement from sequences. In SIGMOD 2005.